\documentclass[aps,prb,showpacs,twocolumn,amsmath,amssymb,superscriptaddress,letterpaper]{revtex4}
\usepackage{times}
\usepackage{amsfonts}
\usepackage{mathrsfs}
\usepackage{graphicx}
\usepackage{dcolumn}
\usepackage{bm}
\usepackage{color}

\usepackage[colorlinks,bookmarks=false,citecolor=blue,linkcolor=red,urlcolor=blue]{hyperref}
\usepackage{color}
\usepackage{siunitx}
\usepackage{bibunits}

\def\be{\begin{equation}}       \def\ee{\end{equation}}
\def\bea{\begin{eqnarray}}      \def\eea{\end{eqnarray}}

\begin{document}

\title{Experimental consequences of $p_z$-wave spin triplet superconductivity in A$_2$Cr$_3$As$_3$}
\author{Xianxin Wu}
\affiliation{ Institute of Physics, Chinese Academy of Sciences,
Beijing 100190, China}

\author{Fan Yang}\email{yangfan\_blg@bit.edu.cn}
\affiliation{School of Physics, Beijing Institute of Technology, Beijing, 100081, China}

\author{Shengshan Qin}
\affiliation{ Institute of Physics, Chinese Academy of Sciences,
Beijing 100190, China}

\author{Heng Fan}  \affiliation{ Institute of Physics, Chinese Academy of Sciences,
Beijing 100190, China} \affiliation{Collaborative Innovation Center of Quantum Matter, Beijing, China}

\author{Jiangping Hu  }\email{jphu@iphy.ac.cn} \affiliation{ Institute of Physics, Chinese Academy of Sciences,
Beijing 100190, China}\affiliation{Department of Physics, Purdue University, West Lafayette, Indiana 47907, USA}
\affiliation{Collaborative Innovation Center of Quantum Matter, Beijing, China}

\date{\today}

\begin{abstract}
The experimental observable properties of the triplet $p_z$-wave pairing state,  proposed by Wu {\em et al.} [arXiv:1503.06707] in quasi-one dimensional A$_2$Cr$_3$As$_3$ materials, are theoretically investigated. This pairing state is characterized by the line nodes on the $k_z=0$ plane on the Fermi surfaces. Based on the three-band tight binding model, we obtain the specific heat, superfluid density, Knight shift and spin relaxation rate and find that all these properties at low temperature ($T\ll T_c$) show powerlaw behaviors and are consistent available experiments. Particularly, the superfluid density determined by the $p_z$-wave pairing state in this quasi-one dimensional system is anisotropic: the in-plane superfluid density varies as $\Delta\rho_{\parallel}\sim T$ but the out-plane one varies as $\Delta\rho_{\perp}\sim T^3$ at low temperature.  The  anisotropic upper critical field reported in experiment is  consistent with the $S_z=0$  (i.e., $(\uparrow\downarrow+\downarrow\uparrow)$)  $p_z$-wave pairing state. We also suggest  the phase-sensitive dc-SQUID measurements to pin down  the triplet $p_z$-wave pairing state.

\end{abstract}


\maketitle

\begin{bibunit}

Recently, superconductivity in chromium-based materials has been revealed, attracting a lot of research interests\cite{Wu2014natcom,Kotegawa2014,Bao2015,Tang2015prb,Tang2015scm}. For CrAs, the superconductivity is achieved by suppressing the magnetic order with pressure\cite{Wu2014natcom,Kotegawa2014}. The close proximity of superconductivity to an helimagnetic order suggests an unconventional pairing mechanism\cite{Wu2014natcom,Kotegawa2014,Shen2014,Kotegawa2015}. The discovery of superconductivity in quasi-one dimensional(Q1D) materials A$_2$Cr$_3$As$_3$(A=K,Rb,Cs)\cite{Bao2015,Tang2015prb,Tang2015scm} is especially interesting due to the rarity of Q1D superconductors. So far, the highest T$_c$$\sim 6.1$K can been achieved in K$_2$Cr$_3$As$_3$ at ambient pressure\cite{Bao2015}. In normal states, these materials possess large electronic specific-heat coefficients\cite{Bao2015} and show non-Fermi liquid transport behavior. Particularly, K$_2$Cr$_3$As$_3$ exhibits the exotic properties of Tomonaga-Luttinger liquids in NMR measurements\cite{Zhi2015}. In superconducting states, many unconventional superconducting properties have been experimentally observed, including linearly temperature dependent penetration depth at $T\ll T_c$\cite{Pang2015}, the absence of Hebel-Slichter coherence peak in $1/T_1$ \cite{Zhi2015} and the extremely large anisotropic upper critical field H$_{c2}(0)$\cite{Bao2015,Tang2015prb,Tang2015scm,Kong2015,Balakirev2015,Wang2015}. All these suggest the existence of line nodes and possible spin triplet pairing in these Q1D superconductors. Furthermore, Raman scattering measurements suggest that electron phonon coupling is rather weak and magnetic fluctuations are coupled to the electronic structure via the lattice\cite{Zhang2015arxiv}.

Theoretical calculations show that $3d$ orbitals of Cr dominate the Fermi surfaces (FSs) consisting of two Q1D $\alpha$ and $\beta$ FSs and one three-dimensional $\gamma$ FS\cite{Jiang2014,Wu2015cpl}. These Q1D materials possess strong frustrated magnetic fluctuations and are nearby a novel in-out co-planar magnetic ground state\cite{Wu2015cpl}. Moreover, the band structure near the Fermi level can be captured by a minimum three-band tight binding model based on the $A'_1(d_{z^2})$ and $E'(d_{xy},d_{x^2-y^2})$ molecular orbitals\cite{Zhou2015,Wu2015rpa,Zhong2015}. In a recent paper\cite{Wu2015rpa}, we adopted combined standard random phase approximation (RPA) approach to study the multi-orbital Hubbard-Hund model in the weak coupling limit and mean-field approach to study the t-J model in the strong coupling limit to investigate the pairing symmetry. Both approaches consistently yield the triplet $p_z$-wave pairing as the leading pairing symmetry for physically realistic parameters. The triplet pairing is driven by the ferromagnetic fluctuations within the sublattice. When considering spin-orbit coupling, the $S_z=0$ component ($\uparrow\downarrow+\downarrow\uparrow$) slightly wins over the $S_z=\pm 1$ ones ($\uparrow\uparrow$,$\downarrow\downarrow$).

In this paper, we investigate the experimental observable properties of the triplet $p_z$-wave pairing state in Q1D A$_2$Cr$_3$As$_3$ materials, characterized by the line nodes on the $k_z=0$ plane on the Fermi surfaces. Based on the three-band tight binding model, we obtain the specific heat, superfluid density, Knight shift and spin relaxation rate and find that all these properties at low temperature show power-law behaviors and are consistent available experiments. To be specific, when $T\ll T_c$, the specific heat varies as $C_v \sim T^2$, the Knight shift along the $\mathbf{d}$-vector ($\parallel \mathbf{z}$ in this system) decays linearly but the other components remain almost unchanged and the spin relaxation rate varies as $1/T_1T \sim T^2$. Particularly, the superfluid density determined by the $p_z$-wave pairing state in this Q1D system is anisotropic: the in-plane superfluid density varies as $\rho_{\parallel}\sim T$ but the out-plane one varies as $\rho_{\perp}\sim T^3$ at low temperature. The phase change along $z$ direction in $p_z$-wave state can be justified in the phase-sensitive measurements. The observed anisotropic upper critical field can be explained by the $p_z$-wave pairing state with $\mathbf{d} \parallel \mathbf{z}$ (i.e. $(\uparrow\downarrow+\downarrow\uparrow)$).

{\bf{\em Model:}} As the states near the Fermi level are contributed by three bands in K$_2$Cr$_3$As$_3$, the minimum model to capture the main physics is given by a three-band tight binding model\cite{Wu2015rpa}. Neglecting spin-orbit coupling (SOC), the tight-binding Hamiltonian is given by,
\begin{equation}
H_{\rm TB}=\sum_{\mathbf{k}}\psi^\dag(\mathbf{k}) h(\mathbf{k}) \psi(\mathbf{k}),\label{TB}
\end{equation}
where $\psi^\dag(\mathbf{k})=[c^\dag_{1\uparrow}(\mathbf{k}),c^\dag_{2\uparrow}(\mathbf{k}),c^\dag_{3\uparrow}(\mathbf{k}),
c^\dag_{1\downarrow}(\mathbf{k}),c^\dag_{2\downarrow}(\mathbf{k}),c^\dag_{3\downarrow}(\mathbf{k})]$. The orbital index $\nu=1,2,3$ represent the $d_{z^2}$ for 1, the $d_{xy}$ for 2, and the $d_{x^2-y^2}$ for 3, respectively. The matrix $h(\mathbf{k})$ has been given in Ref.\onlinecite{Wu2015rpa}. When considering the superconducting pair, the mean field Hamiltonian can be written as,
\begin{eqnarray}
H_{SC}&=&\frac{1}{2}\sum_{\mathbf{k}}\Psi^\dag(\mathbf{k}) h_{SC}(\mathbf{k}) \Psi(\mathbf{k}),\nonumber\\
h_{SC}(\mathbf{k})&=&\left(\begin{array}{cc}
h(\mathbf{k}) & \Delta(\mathbf{k}) \\
 \Delta^{\dagger}(\mathbf{k}) & -h^{\star}(-\mathbf{k}) \\
 \end{array}\right),\label{Hamiltonian}
\end{eqnarray}
where $\Psi^\dag(\mathbf{k})=[\psi^\dag(\mathbf{k}),\psi^T(-\mathbf{k})]$. Here we only consider the intraorbital pair, the pairing term $\Delta(\mathbf{k})$ is given by,
\begin{eqnarray}
\Delta(\mathbf{k})&=&\left(\begin{array}{cccccc}
 \Delta^{\uparrow\uparrow}(\mathbf{k}) & \Delta^{\uparrow\downarrow}(\mathbf{k}) \\
 \Delta^{\downarrow\uparrow}(\mathbf{k}) & \Delta^{\downarrow\downarrow}(\mathbf{k}) \\
  \end{array}\right),\\
\Delta^{\sigma\sigma'}(\mathbf{k})&=&\left(\begin{array}{ccc}
 \Delta^{\sigma\sigma'}_{d_{z^2}}(\mathbf{k}) & & \\
 & \Delta^{\sigma\sigma'}_{d_{x^2-y^2}}(\mathbf{k}) &\\
 &  & \Delta^{\sigma\sigma'}_{d_{xy}}(\mathbf{k})\\
 \end{array}\right).
\end{eqnarray}
In the following, unless otherwise specified, we shall consider the $p_z$-wave spin triplet state with $\mathbf{d}=\Delta sink_z \mathbf{z}$, which means $ \Delta^{\uparrow\uparrow}(\mathbf{k})=\Delta^{\downarrow\downarrow}(\mathbf{k})=0$,$\Delta_{\nu}^{\uparrow\downarrow}(\mathbf{k})
=\Delta_{\nu}^{\downarrow\uparrow}(\mathbf{k})=\Delta_{\nu}sink_z$. Due to the existence of line nodes in the $k_z=0$ plane, the quasi-particles exhibit linear density of states at low energies\cite{Wu2015rpa}.

Diagonalizing the BdG matrix $h_{SC}(\mathbf{k})$ in Eq.(\ref{Hamiltonian}), we obtain the eigenvalue $E_{\mathbf{k}n}$ as the band energy and the eigenvector $\phi_{n\mathbf{k}}$ as the eigen state. In the following calculations at finite temperature $T$, the $T-$dependence
of the gap amplitudes is assumed as $\Delta_{\nu}(t)=\Delta^0_{\nu}\delta(t)$. Here $\Delta^0_{x^2-y^2}=\Delta^0_{xy}=4\Delta^0_{z^2}=0.83~meV$ are obtained in Ref.\onlinecite{Wu2015rpa}, and $\delta(t)$ is the normalized BCS gap at the reduced temperature $t=T/T_c$\cite{M¨¹hlschlegel1959}. We have also performed calculations using the temperature dependent gap from solving the t-J model and found similar temperature-dependence of the following experimental observables at sufficiently low temperature $T\ll T_c$.

\begin{figure}[tb]
\centerline{\includegraphics[height=6.5 cm]{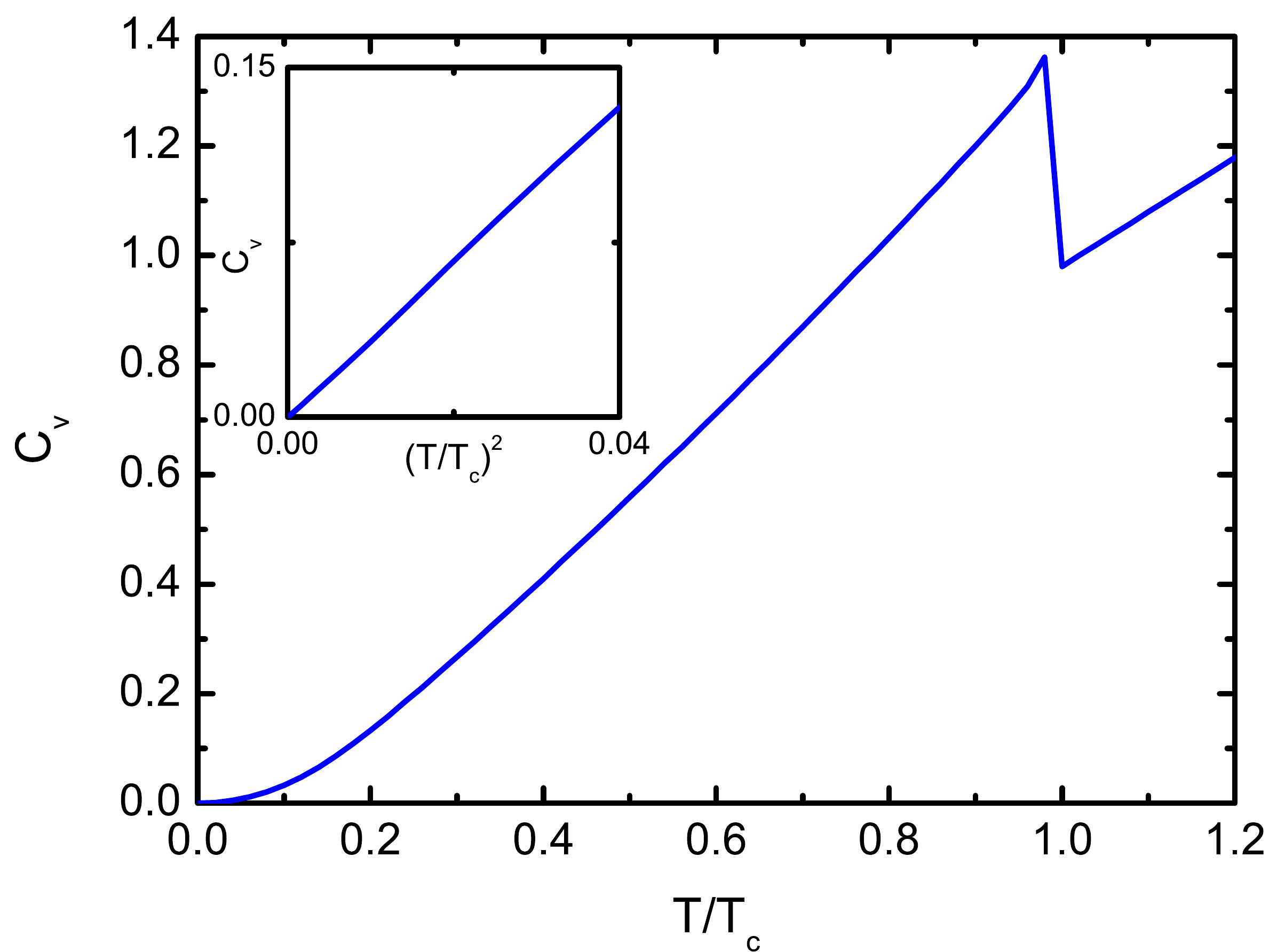}}
\caption{(color online) The electronic specific heat $C_v$ as a function of temperature for the $p_z$-wave state. The inset shows the specific heat at low temperature.  \label{specificheat} }
\end{figure}

{\bf{\em Specific heat:}} The specific heat of this system is given by,
\begin{eqnarray}
C_v=\frac{1}{2N}\sum_{\mathbf{k}n}E_{\mathbf{k}n}\frac{dn_F(E_{\mathbf{k}n})}{dT},
\end{eqnarray}
where $n_F(E_{\mathbf{k}n})=\frac{1}{e^{E_{\mathbf{k}n}/k_BT}+1}$ is the Fermi-Dirac distribution function. Here $\frac{dn_F(E_{\mathbf{k}n})}{dT}$ can be further evaluated as,
\begin{eqnarray}
\frac{dn_F}{dT}=n_E(E_{\mathbf{k}n})[1-n_E(E_{\mathbf{k}n})](\frac{E_{\mathbf{k}n}}{k_BT^2}-\frac{1}{k_BT}\frac{dE_{\mathbf{k}n}}{dT}).
\end{eqnarray}
 The calculated specific heat $C_v$ is shown in Fig.\ref{specificheat}. Near $T_c$, $C_v$ jumps, reflecting the superconducting phase transition. At low temperatures, $C_v\propto T^2$ (inset in Fig.\ref{specificheat}), reflecting the line gap nodes in the $p_z$-wave state, consistent with experiments\cite{Luo2015}.

{\bf{\em Superfluid density:}} The temperature-dependence of the superfluid density of the system takes on conflicting characteristics for different available experiments. On the one hand, through a tunnel diode oscillator, measurements of the temperature dependent penetration depth were done in K$_2$Cr$_3$As$_3$ and a linear relationship was found at low temperature, which suggests the existence of line nodes in superconducting gap\cite{Pang2015}. On the other hand, the temperature dependence of the superfluid density obtained from the  muon-spin relaxation measurements fits well
to an isotropic $s$-wave character for the superconducting gap\cite{Adroja2015}. However, from our point of view, such conflicting experimental behaviors may be attributed to the anisotropic superfluid density of the Q1D $p_z$-wave pairing state here. Actually, our results suggest that at low temperature, while the in-plane superfluid density scales with $T$, the out-plane one scales with $T^3$ (which cannot be easily distinguished from exponential function for $T/T_c\ll 1$).

The linear response of the system to an external magnetic field is detailed in the supplementary materials. In the superconducting state, the superfluid density $\rho$ is proportional to the response kernel $K(q\rightarrow 0,\omega=0)$, with $\rho=\rho_p+\rho_d$ and $K(q,\omega)=K^p(q,\omega)+K^d(q,\omega)$. In A$_2$Cr$_3$As$_3$, the paramagnetic part $\rho_p$ and diamagnetic part $\rho_d$ read,
\begin{eqnarray}
\rho^{ss}_p&=&\frac{1}{4N}\sum_{\mathbf{k}mn}|\phi^\dag_{m\mathbf{k}}F^p_{s}(\mathbf{k})\phi_{n\mathbf{k}}|^2\frac{n_F(E_{\mathbf{k}m}) -n_F(E_{\mathbf{k}n})}{E_{\mathbf{k}m}-E_{\mathbf{k}n}},\nonumber\\
\rho^{ss}_d&=&\frac{1}{4N}\sum_{\mathbf{k}n}|\phi^\dag_{n\mathbf{k}}F^d_{ss}(\mathbf{k})\phi_{n\mathbf{k}}|^2n_F(E_{\mathbf{k}n}).
\end{eqnarray}
Here $s=x,y,z$ and $F^p_{s}(\mathbf{k})$ and $F^d_{ss}(\mathbf{k})$ are given by,
\begin{eqnarray}
F^p_{s}(\mathbf{k})&=&\left(\begin{array}{cc}
f^p_s(\mathbf{k}) & 0 \\
0 & -f^{p\star}_s(-\mathbf{k}) \\
 \end{array}\right), \\
 F^d_{ss}(\mathbf{k})&=&\left(\begin{array}{cc}
f^d_{ss}(\mathbf{k}) & 0 \\
0 & -f^{d\star}_{ss}(-\mathbf{k}) \\
 \end{array}\right),
\end{eqnarray}
where $f^p_s(\mathbf{k})$ and $f^d_{ss}(\mathbf{k})$ are given in the supplementary materials. For the intraband contribution, the Linhard function should be replaced with $\frac{\partial n_F(E_{\mathbf{k}m})}{\partial E_{\mathbf{k}m}}$.

\begin{figure}[tb]
\centerline{\includegraphics[height=6.5 cm]{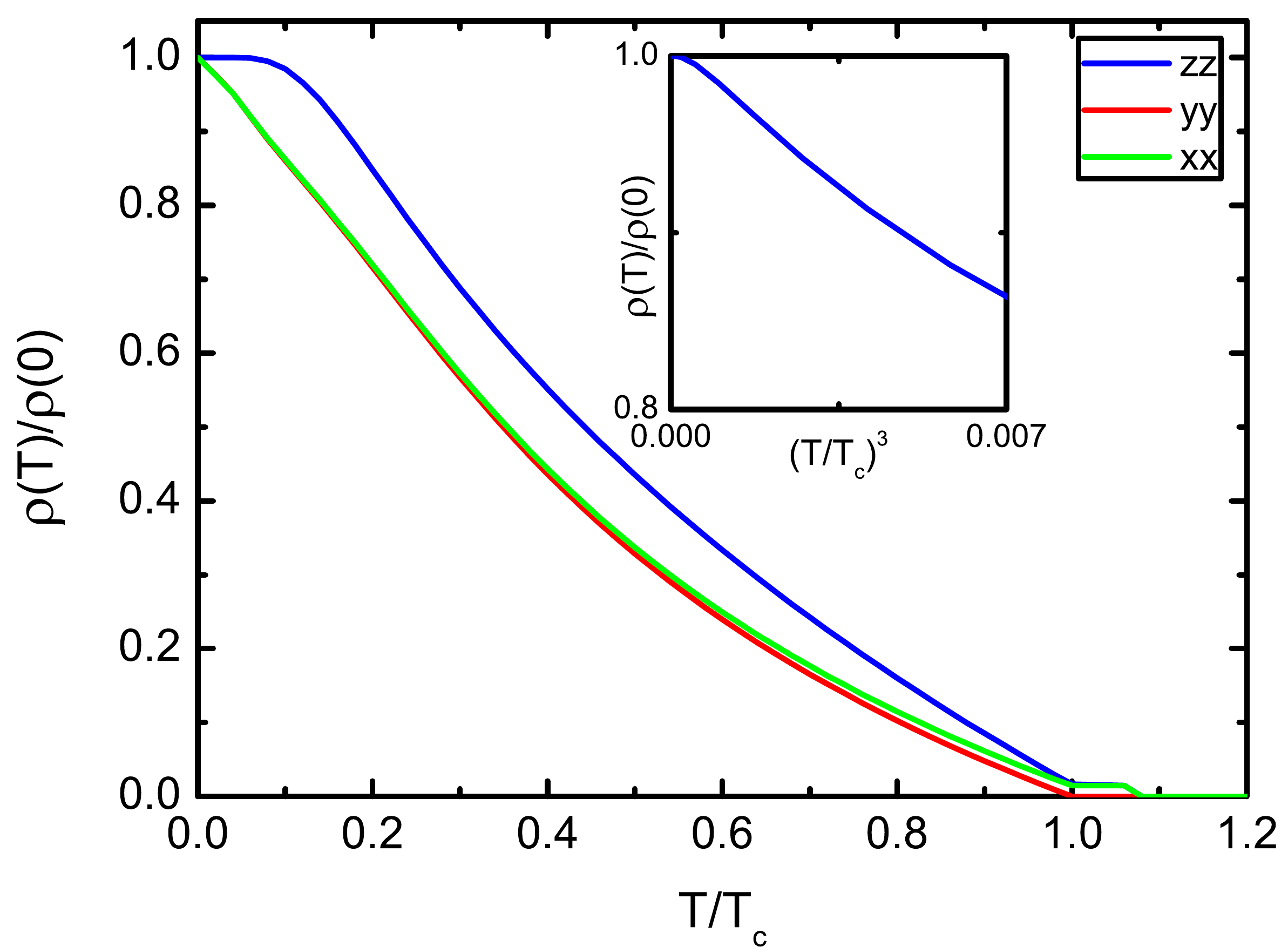}}
\caption{(color online) The normalized superfluid density $\rho^{ss}$($s=x,y,z$) as a function of temperature for the $p_z$-wave state. Due to the quasi-one dimensionality of the system, the absolute superfluid density $\rho^{zz}(0):\rho^{xx/yy}(0)\approx 20$.  The inset shows the superfluid density at low temperature.  \label{superfluid} }
\end{figure}
The temperature dependent normalized superfluid density $\rho$ is shown in Fig.\ref{superfluid}. We find that $\rho$ is anisotropic, which is caused by the $p_z$-wave pairing state in this Q1D system. Firstly, the superfluid density along the $z$-axis is much larger than that in the $xy$-plane due to the Q1D characteristic of the systems. Secondly, while the superfluid density $\Delta\rho=\rho(T)-\rho(0)$ in the $xy$-plane varies as $\Delta\rho_{\parallel}\sim T$, the out-plane superfluid density $\Delta\rho_{\perp}$ varies slowly at low temperature, which seems at first glance like exponentially. However, further calculations show that the out-plane superfluid density varies as $\Delta\rho_{\perp}\sim T^3$, which cannot be easily distinguished from the exponential function at low $T$. To understand the qualitatively different behaviors of $\rho_{\parallel}$ and $\rho_{\perp}$ at low temperatures, we have made a thorough investigation in the Supplementary Information. Here, we underdraw the physics. As the diamagnetic part $\rho_d$ is almost temperature independent, we can focus on the paramagnetic part $\rho_{p}$, which represents for the consuming of the superfluid density by the nodal quasi-particle excitations aroused by the paramagnetic current operator $J^p$. At low temperatures, these excitations mainly take place near $k_z=0$ for the $p_z$-wave pairing. For these small $k_z$, the in-plane current is $J^p_{x/y}\sim g_1(k_x,k_y)(1-\frac{1}{2}k^2_z)$, which leads to $\Delta\rho_{\parallel}\sim T$. However, the out-plane current is $J^p_{\perp}\sim k_zg_2(k_x,k_y)$, which goes to zero for small $k_z$ and the aroused quasi-particle excitations are strongly suppressed, which leads to $\Delta\rho_{\perp}\sim T^3$. Such low-temperature power-law behaviors of the superfluid density are the consequences of the polar states with an equatorial line of nodes\cite{Einzel1986,Gross1986}. The obtained anisotropic superfluid density for $p_z$-wave state seems to be consistent with experiments, where both linear and seemingly exponential temperature dependence of superfluid density were observed.

{\bf{\em NMR:}} Recently, the nuclear magnetic resonance(NMR) measurements suggest unconventional nature of superconductivity, reflected in the absence of the Hebel-Slichter coherence peak below T$_c$\cite{Zhi2015}. Moreover, the temperature dependence of $1/T_1$ below T$_c$ shows power-law behavior. Now we investigate the spin-relaxation rate $1/T_1$, as well as the Knight shift $K$ for the $p_z$-wave triplet pairing state.

In general the spin susceptibility is defined as,
\begin{eqnarray}
\chi^{st}_{\nu\mu}(q,i\omega_n)&=&\int^{\beta}_{0}\langle T_{\tau} S^{s}_{\nu}(q,\tau) S^{t}_{\mu}(-q,0)  \rangle e^{i\omega_n\tau} d\tau,
\end{eqnarray}
where $S^t_{\mu}$ is the $t$ component of spin operator for orbital $\mu$. The NMR spin-relaxation rate reads,
\begin{eqnarray}
\frac{1}{T_1T}&\varpropto&\lim\limits_{\omega \to 0 }\sum_{\mathbf{q},s,\mu,\nu}|A(\mathbf{q})|^2\frac{Im\chi^{ss}_{\mu\nu}(\mathbf{q},\omega+i0^+)}{\omega}\nonumber\\
&=& -\frac{1}{4N^2}\sum_{\mathbf{k}\mathbf{k}',mns}A(\mathbf{k}'-\mathbf{k})|\phi^\dag_{m\mathbf{k}}S_s\phi_{n\mathbf{k}'}|^2 \nonumber \\
&&\frac{\partial n_F(E)}{\partial E}|_{E=E_{\mathbf{k}m}}\delta(E_{\mathbf{k}m}-E_{\mathbf{k}'n}),
\end{eqnarray}
where the geometrical structure factor $A(\mathbf{q})$ has been set to be 1 in our calculation for simplicity. Here the spin matrices are $S_{x,z}=\eta_3\otimes\sigma_{x,z}\otimes I_0$ and $S_{y}=\eta_0\otimes\sigma_{y}\otimes I_0$, where the Pauli matrices $\sigma$ and $\eta$ act in the spin and Nambu space and the $3\times3$ identity $I_0$ acts in the orbital space.

\begin{figure}[tb]
\centerline{\includegraphics[height=6.5 cm]{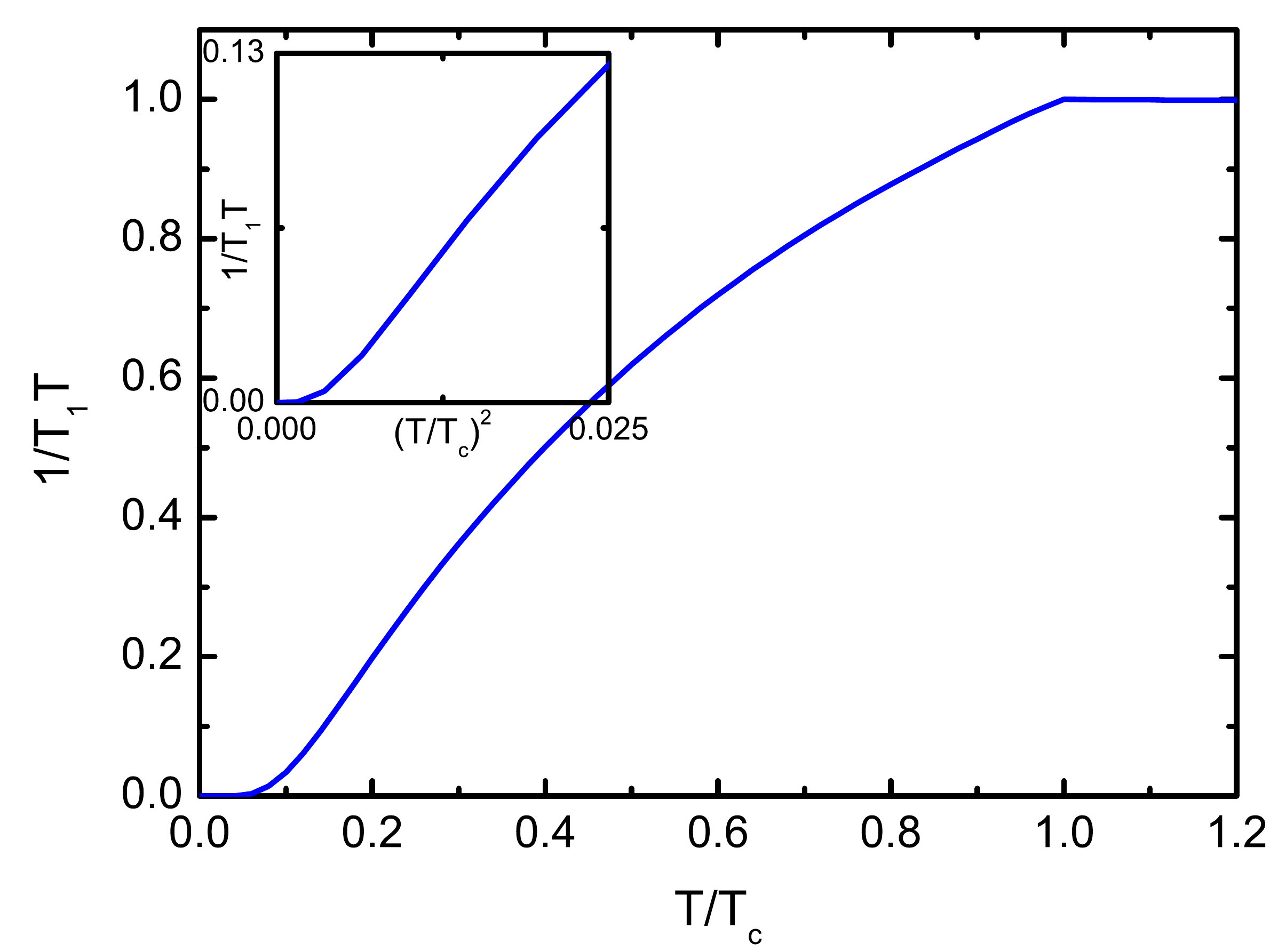}}
\caption{(color online) The spin relaxation rate $1/T_1T$ as a function of $\frac{T}{T_c}$ for the $p_z$-wave state. Inset shows $1/T_1T$ at low temperature.  \label{NMR} }
\end{figure}

Fig.\ref{NMR} shows the spin relaxation rate $1/T_1T$ for the $p_z$-wave state. The most distinctive feature of $1/T_1T$ is that there is no Hebel-Slichter coherence peak below T$_c$. At low temperature, $1/T_1T$ shows powerlaw behavior, i.e., $1/T_1\sim T^3$, because of the line nodes in superconducting gap. Such results are consistent with the experimental data although the power law exponent is slightly different from the experimental one\cite{Zhi2015}.

\begin{figure}[t]
\centerline{\includegraphics[height=5.5 cm]{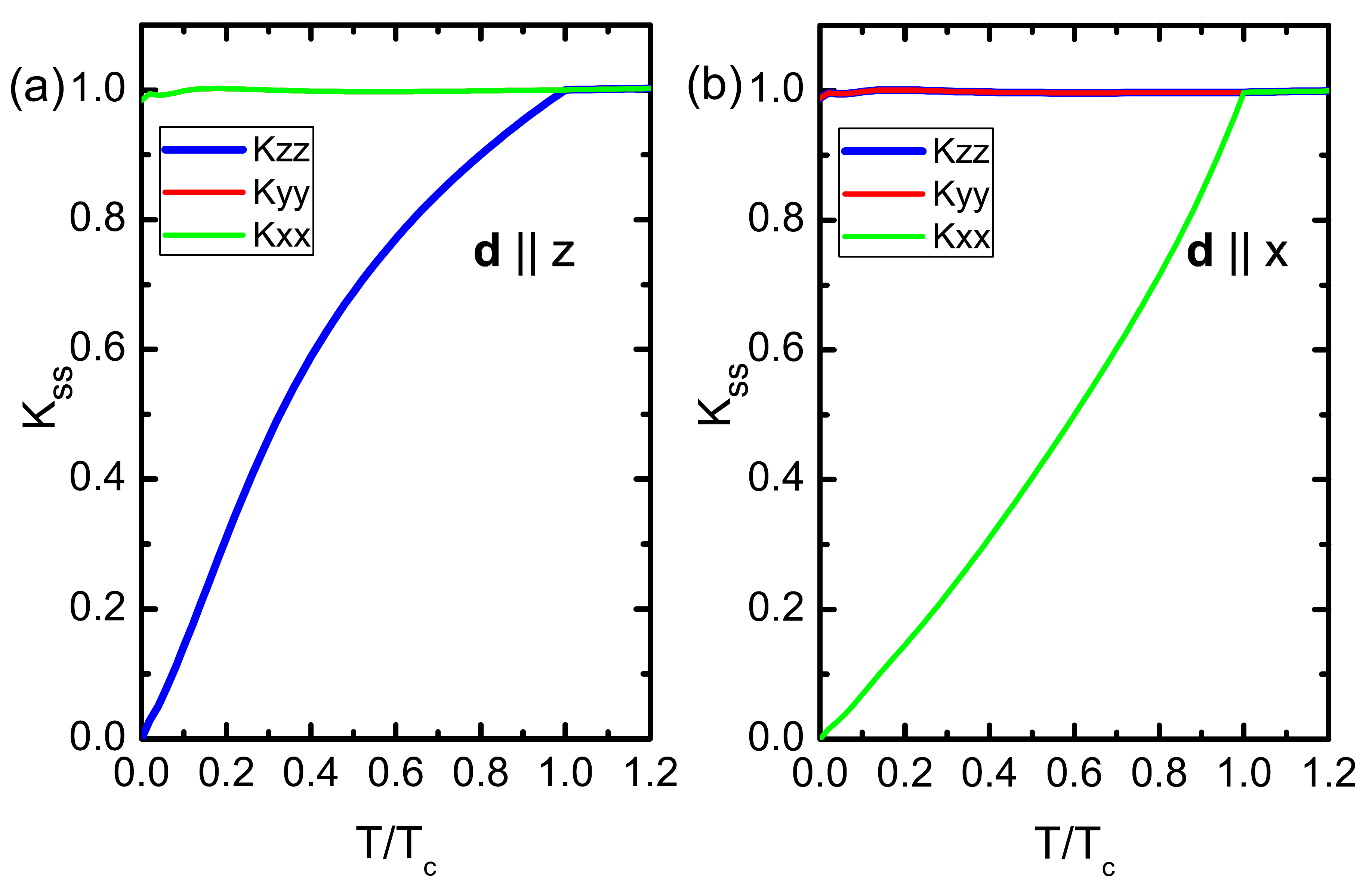}}
\caption{(color online) The Knight shift $K_{ss}$($s=x,y,z$) as a function of temperature for the $p_z$-wave state. (a) $\mathbf{d}=\Delta sink_z \mathbf{z}$ and the corresponding parameter is $\Delta_{\nu}^{\uparrow\downarrow}=\Delta_{\nu}^{\downarrow\uparrow}$, (b) $\mathbf{d}=\Delta sink_z \mathbf{x}$ and the corresponding parameter is $\Delta_{\nu}^{\uparrow\uparrow}=-\Delta_{\nu}^{\downarrow\downarrow}$.  \label{knightshift} }
\end{figure}

The Knight shift reads,
\begin{eqnarray}
K_{ss}&\varpropto& \sum_{\mu\nu}\chi^{ss}_{\mu\nu}(0,0) \nonumber\\
&=&-\frac{1}{4N}\sum_{\mathbf{k}mn}|\phi^\dag_{m\mathbf{k}}S_s\phi_{n\mathbf{k}}|^2\frac{n_F(E_{\mathbf{k}m}) -n_F(E_{\mathbf{k}n})}{E_{\mathbf{k}m}-E_{\mathbf{k}n}}.\nonumber\\
\end{eqnarray}
The Knight shift experiment can be used to distinguish between spin-singlet and spin-triplet pairings. Further more, for the triplet pairing, it can identify the multi-component order parameter through the $\mathbf{d}$-vector defined as,
 \begin{eqnarray} \mathbf{d}_{\mathbf{k}}=[\frac{\Delta_{\mathbf{k}\downarrow\downarrow}-\Delta_{\mathbf{k}\uparrow\uparrow}}{2},-i\frac{\Delta_{\mathbf{k}\downarrow\downarrow}+\Delta_{\mathbf{k}\uparrow\uparrow}}{2},
 \Delta_{\mathbf{k}\uparrow\downarrow}].
 \end{eqnarray}

The obtained Knight shifts $K_{ss}$ for the $\mathbf{d}=\Delta sink_z \mathbf{z}$ and $\mathbf{d}=\Delta sink_z \mathbf{x}$ pairing states are shown in Fig.\ref{knightshift}(a) and (b), respectively. In the superconducting state, the component of Knight shift along $\mathbf{d}$ is strongly suppressed while the other components almost unchange with the decreasing of temperature. For the $\mathbf{d}=\Delta sink_z \mathbf{z}$ state obtained in Ref.\onlinecite{Wu2015rpa}, the spin orientation is confined in the $xy$-plane. Therefore, the in-plane magnetic field can polarize the spin without breaking the Cooper pair. On the contrary, the out-plane magnetic field need to break the Cooper pairing in order to cause spin splitting. Therefore, this excitation is linearly suppressed at low temperature due to the existence of line nodes in superconducting state. Similarly, we can understand the Knight shift for the $\mathbf{d}=\Delta sink_z \mathbf{x}$ state.


\begin{figure}[tb]
\centerline{\includegraphics[height=10 cm]{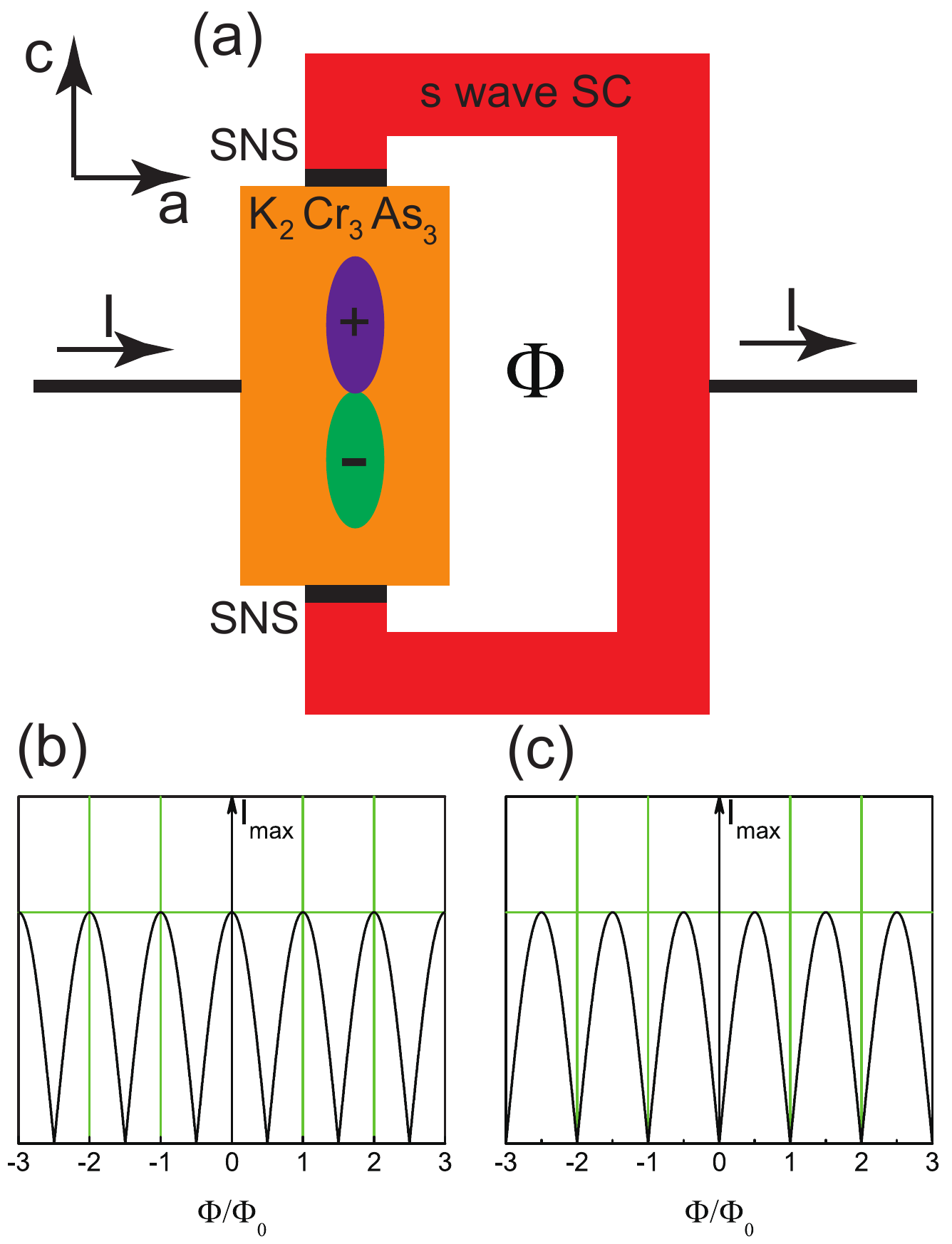}}
\caption{(color online) Experimental geometry for a SQUID phase sensitive probe and interference patterns of a SQUID. (b) s-wave case. (c)$p_z$-wave case. \label{josephsonjunction} }
\end{figure}

{\bf{\em Phase sensitive measurements:}} The triplet $p_z$-wave pairing predicted here can be detected by the dc SQUID, a phase-sensitive device which has been adopted in determining the pairing symmetries of such superconducting systems as the cuprates \cite{squid,Wollman1993,Mathai1995}, the Sr$_2$RuO$_4$ \cite{josephson} and others\cite{yang}. The proposed configuration is shown in Fig.\ref{josephsonjunction}(a), where two superconductor-normal metal-superconductor (SNS) Josephson tunneling junctions are formed on the two opposite edges in the $z$ direction of K$_2$Cr$_3$As$_3$, which are connected by a loop of a conventional $s$-wave superconductor, forming a bimetallic ring with a magnetic flux $\Phi$ threading through the loop.

As a result of the interference between the two branches of Josephson supercurrent, the maximum total supercurrent (the critical current) $I_c$ in the circuit modulates with $\Phi$ according to
\begin{align}
I_c(\Phi)=2I_{0}\left|\cos\left(\pi\frac{\Phi}{\Phi_0}
+\frac{\phi_0}{2}\right)\right|.
\end{align}
Here $I_0$ is the critical current of one Josephson junction, $\Phi_{0}=h/2e$ is the basic flux quantum, $\phi_0$ is the phase shift along $z$ direction of K$_2$Cr$_3$As$_3$.

If K$_2$Cr$_3$As$_3$ is a conventional superconductor, there is no phase change along $z$ direction so that $\phi_0=0$ and the maximum current should show a maximum for $\Phi=n\Phi_0$, as shown in Fig.\ref{josephsonjunction}(b). However, for $p_z$-wave state there a $\pi$ phase shift along $z$ direction, that is, $\phi_0=\pi$. The corresponding current should show a maximum for $\Phi=(n+\frac{1}{2})\Phi_0$, as shown in Fig.\ref{josephsonjunction}(c). The phase change obtained in this experiment can provide a verdict for the $p_z$-wave pairing state.

{\bf{\em Upper critical magnetic field:}} One of the most striking features of A$_2$Cr$_3$As$_3$ superconductors is that the upper critical field $H_{c2}$ is very high and severely exceeds the Pauli limit\cite{Bao2015,Tang2015prb,Tang2015scm,Kong2015,Balakirev2015,Wang2015}. Furthermore, the $H_{c2}$ along the chains shows paramagnetic-limited behavior but the $H_{c2}$  perpendicular to the chains does not, which results in a smaller $H_{c2}$ along $z$ axis at 0 K. The observed high $H_{c2}$ can hardly be attributed to the spin orbital effect but may suggest triplet pairing. Actually, the high anisotropic $H_{c2}$ observed in these materials is consistent with the proposed $p_z$-wave spin triplet state with $\mathbf{d}=\Delta sink_z \mathbf{z}$ (i.e. $(\uparrow\downarrow+\downarrow\uparrow)$). The magnetic field parallel to $\mathbf{d}$($z$ direction) will cause Zeeman splitting and break the Cooper pair, which leads to the paramagnetic-limited behavior. However, the in-plane magnetic field will not cause pair breaking because this pairing can be equally looked upon as the iteration of two equal spin pairings with opposite spin directions in the $x,y$-plane. Thus, there is no paramagnetic suppression of superconductivity for in-plane magnetic field. The extraordinary crossover of $H_{c2}(T)$ curves in K$_2$Cr$_3$As$_3$ is rather similar to that of a heavy fermion system UPt$_3$\cite{Shivaram1986}. The  crossover in the latter can hardly be explained by a spin singlet state but can be well explained by a spin triplet state with SOC\cite{Choi1991,Choi1993}. Therefore, the crossover in K$_2$Cr$_3$As$_3$ may be similarly understood to be caused by the $p_z$-wave spin triplet state with SOC. Further detailed calculations are needed.

In conclusion, we have investigated the experimental consequences of the $p_z$-wave spin triplet superconductivity in A$_2$Cr$_3$As$_3$. Based on the three-band tight binding model, we obtain the specific heat, superfluid density, Knight shift and spin relaxation rate and find that all these properties at low temperature show powlaw temperature-dependences and are consistent with available experiments. Due to the existence of line nodes in $k_z=0$ plane, the quasi-particles exhibit linear density of states at low energies. The specific heat exhibits a quadratic temperature dependence at $T\ll T_c$. The $p_z$-wave pairing state leads to an anisotropic superfluid density in the Q1D materials: the in-plane superfluid density varies as $\Delta\rho_{\parallel}\sim T$ but the out-plane one varies as $\Delta\rho_{\perp}\sim T^3$ at low temperature, which is consistent with experimental data. In the spin relaxation rate, there is no Hebel-Slichter coherence peak below T$_c$ and $1/T_1$ varies as $T^3$ at low temperature. In our obtained superconducting state with $\mathbf{d}=\Delta sink_z \mathbf{z}$ (which means a $(\uparrow\downarrow+\downarrow\uparrow)$ pairing), the out-plane Knight shift is linearly suppressed at low temperature but the in-plane one remains almost unchanged. The phase structure in the $p_z$-wave state can be justified in the phase-sensitive dc-SQUID measurements. Finally, the experimentally observed anisotropic upper critical field can be explained by the $p_z$-wave pairing state with $\mathbf{d} \parallel \mathbf{z}$ obtained here.

{\bf Acknowledgments}: This work is supported in part by  MOST of China (2012CB821400,2011CBA00100,2015CB921300), NSFC(11190020,91221303,11334012,11274041) and  ``Strategic Priority Research Program (B)" of the Chinese Academy of Sciences( XDB07020200). F.Y is also  supported by the NCET program under Grant No. NCET-12-0038.

\end{bibunit}

\begin{bibunit}

\clearpage
\pagebreak
\onecolumngrid
\widetext
\begin{center}
\textbf{\large Supplementary materials for ``Experimental consequences of $p_z$-wave spin triplet superconductivity in A$_2$Cr$_3$As$_3$''}
\end{center}

\setcounter{equation}{0}
\setcounter{figure}{0}
\setcounter{table}{0}
\makeatletter
\renewcommand{\theequation}{S\arabic{equation}}
\renewcommand{\thefigure}{S\arabic{figure}}
\renewcommand{\bibnumfmt}[1]{[S#1]}
\renewcommand{\citenumfont}[1]{S#1}

\appendix

\subsection{The linear response of the system to an external magnetic field}

In the presence of a weak external magnetic field taken as a perturbation, the coupling of the system to this field can be introduced via the Peierls substitution $c^{\dag}_{i\alpha\sigma}c_{j\beta\sigma}\rightarrow c^{\dag}_{i\alpha\sigma}e^{i\int^{i}_{j}\bf{A}\cdot d\bf{l}}c_{j\beta\sigma}$ in the tight-binding Hamiltonian (\ref{TB}), where $\textbf{A}$ is the vector potential. This Hamiltonian can be expanded up to the second-order terms with respect to $\textbf{A}$ as
\begin{eqnarray}
H(\textbf{A})=H(0)-\sum_{i}[\mathbf{J}^p(i)\mathbf{A}(i)-\frac{1}{2}K^d(i)\mathbf{A}^2(i)],
\end{eqnarray}
where $\mathbf{J}^p$ is the paramagnetic current.
Then the current $J_{s}(i)$ $(s=x,y,z)$ is given by,
\begin{eqnarray}
J_s(i)&=&-\frac{\delta H(\mathbf{A})}{\delta A_s(i)}=J^p_s(i)-K^d_{st}(i)A_t(i).
\end{eqnarray}
In momentum space, the total current can be further written as a response of the system to the vector potential,
\begin{eqnarray}
J_s(q)=-\sum_{t}[K^p_{st}(q)+K^d_{st}(q)]A_t(q).
\end{eqnarray}

Here $K^p_{st}$ and $K^d_{st}$ are the paramagnetic and diamagnetic response kernels, respectively.  The diamagnetic response kernel in this system reads,
\begin{eqnarray}
K^d_{st}(q)=\sum_{\mathbf{k}}\langle \psi^\dag(\mathbf{k}) f^d_{st}(\mathbf{k+q}) \psi(\mathbf{k+q}) \rangle,
\end{eqnarray}
with $f^d_{st}(\mathbf{k+q})=\frac{\partial^2h(k)}{\partial k_s\partial k_t}|_{\mathbf{k}=\mathbf{k}+\mathbf{q}}$. The paramagnetic response kernel is obtained through the current-current correlation function in linear response,
\begin{eqnarray}
K^p_{st}(q,\omega)&=&-i\int^{\infty}_{-\infty}e^{i\omega (t-t')}\theta(t-t')\langle [J^p_s(q,t),J^p_t(-q,t')]\rangle dt,
\end{eqnarray}
where $J^p_s(q)=\sum_{\mathbf{k}}\psi^\dag(\mathbf{k}) f^p_{s}(\mathbf{k+q}) \psi(\mathbf{k+q})$ and $f^p_{s}(\mathbf{k+q})=\frac{\partial h(k)}{\partial k_s}|_{\mathbf{k}=\mathbf{k}+\mathbf{q}}$. The corresponding Matsubara Green function is,
\begin{eqnarray}
K^p_{st}(q,i\omega_n)=\int^{\beta}_{0}\langle T_{\tau} J^p_s(q,\tau) J^p_t(-q,0)   \rangle e^{i\tau\omega_n}d\tau
\end{eqnarray}
with $\omega_n=\frac{2n\pi}T$. The retarded Green function can be obtained upon analytic continuation, $K^p_{st}(q,\omega)=K^p_{st}(q,i\omega_n\rightarrow \omega+i\delta)$.

\subsection{The current operators}

The paramagnetic current operator is given by $J^p_s(q)=\sum_{\mathbf{k}}\psi^\dag(\mathbf{k}) f^p_{s}(\mathbf{k+q}) \psi(\mathbf{k+q})$ and the diamagnetic response kernel is $K^d_{st}(q)=\sum_{\mathbf{k}}\langle \psi^\dag(\mathbf{k}) f^d_{st}(\mathbf{k+q}) \psi(\mathbf{k+q}) \rangle$. In the following, we provide the explicit formula of the above $f^p_{s}$ and $f^d_{st}$ matrices.

Let $x=\frac{\sqrt{3}}{2}{k_xa_0}$, $y=\frac{1}{2}k_ya_0$ and $z=\frac{1}{2}k_zc_0$, the matrix elements $f^p_{z}(\mathbf{k})$ and $f^d_{zz}(\mathbf{k})$ are,
\begin{eqnarray}
f^p_{z11}&=&-2c_0s^{11}_{yz}{\rm sin} 2z(2{\rm cos}2y+4{\rm cos} x {\rm cos} y )-2c_0s^{11}_{z2}sin2z-4c_0s^{11}_{z4}sin4z-6c_0s^{11}_{z6}sin6z,\nonumber\\
f^p_{z22}&=&-2c_0s^{22}_{z2}sin2z-4c_0s^{22}_{z4}sin4z\nonumber\\
f^p_{z33}&=&f^p_{z22},f^p_{z12}=f^p_{z13}=f^p_{z23}=0,\label{fpz}
\end{eqnarray}
\begin{eqnarray}
f^{d}_{zz11}&=&-2c^2_0s^{11}_{yz}{\rm cos} 2z(2{\rm cos}2y+4{\rm cos} x {\rm cos} y )-2c^2_0s^{11}_{z2}cos2z-8c^2_0s^{11}_{z4}cos4z-18c^2_0s^{11}_{z6}cos6z,\nonumber\\
f^{d}_{zz22}&=&-2c^2_0s^{22}_{z2}cos2z-8c^2_0s^{22}_{z4}cos4z\nonumber\\
f^{d}_{zz33}&=&f^{d}_{zz22},f^{d}_{zz12}=f^{d}_{zz13}=f^{d}_{zz23}=0
\end{eqnarray}
The matrix elements $f^p_{y}(\mathbf{k})$ and $f^d_{yy}(\mathbf{k})$ are,
\begin{eqnarray}
f^p_{y11}&=&-2a_0(s^{11}_{yz}+2s^{11}_{yz}cos2z)({\rm sin}2y+{\rm cos} x {\rm sin} y )\nonumber\\
f^p_{y12}&=&2ia_0s^{12}_y{\rm cos}2y+ia_0s^{12}_y{\rm cos}y{\rm cos}x+\sqrt{3}a_0s^{12}_{2y}{\rm cos}y{\rm sin}x\nonumber\\
f^p_{y13}&=&-2a_0s^{12}_{2y}{\rm sin}2y+i\sqrt{3}a_0s^{12}_{1y}{\rm sin}y{\rm sin}x+a_0s^{12}_{2y}{\rm sin}y {\rm cos}x \nonumber\\
f^{p}_{y22}&=&-2a_0s^{22}_{11y}{\rm sin}2y-\frac{1}{2}a_0(s^{22}_{11y}+3s^{22}_{22y}){\rm cos}x {\rm sin}y\nonumber\\
f^{p}_{y23}&=&\frac{\sqrt{3}}{2}a_0(s^{22}_{11y}-s^{22}_{22y}){\rm sin}x {\rm cos}y+2ia_0s^{22}_{12y}{\rm cos}2y-2ia_0s^{22}_{12y}{\rm cos}x{\rm cos}y\nonumber\\
f^{p}_{y33}&=&-2a_0s^{22}_{11y}{\rm sin}2y-\frac{1}{2}a_0(3s^{22}_{11y}+s^{22}_{22y}){\rm cos}x {\rm sin}y,\label{fpy}
\end{eqnarray}
\begin{eqnarray}
f^{d}_{yy11}&=&-2a^2_0(s^{11}_{yz}+2s^{11}_{yz}cos2z)({\rm cos}2y+\frac{1}{2}{\rm cos} x {\rm cos} y )\nonumber\\
f^{d}_{yy12}&=&-2ia^2_0s^{12}_y{\rm sin}2y-\frac{1}{2}ia^2_0s^{12}_y{\rm sin}y{\rm cos}x-\frac{\sqrt{3}}{2}a^2_0s^{12}_{2y}{\rm sin}y{\rm sin}x\nonumber\\
f^{d}_{yy13}&=&-2a^2_0s^{12}_{2y}{\rm cos}2y+i\frac{\sqrt{3}}{2}a^2_0s^{12}_{1y}{\rm cos}y{\rm sin}x+\frac{1}{2}a^2_0s^{12}_{2y}{\rm cos}y {\rm cos}x \nonumber\\
f^{d}_{yy22}&=&-2a^2_0s^{22}_{11y}{\rm cos}2y-\frac{1}{4}a^2_0(s^{22}_{11y}+3s^{22}_{22y}){\rm cos}x {\rm cos}y\nonumber\\
f^{d}_{yy23}&=&-\frac{\sqrt{3}}{4}a^2_0(s^{22}_{11y}-s^{22}_{22y}){\rm sin}x {\rm sin}y-2ia^2_0s^{22}_{12y}{\rm sin}2y+ia_0s^{22}_{12y}{\rm cos}x{\rm sin}y\nonumber\\
f^{d}_{yy33}&=&-2a^2_0s^{22}_{11y}{\rm cos}2y-\frac{1}{4}a^2_0(3s^{22}_{11y}+s^{22}_{22y}){\rm cos}x {\rm cos}y
\end{eqnarray}

The matrix elements $f^p_{x}(\mathbf{k})$ and $f^d_{xx}(\mathbf{k})$ are,
\begin{eqnarray}
f^{p}_{x11}&=&-2\sqrt{3}a_0(s^{11}_y+2s^{11}_{yz}{\rm cos}2z){\rm sin}x{\rm cos}y \nonumber\\
f^{p}_{x12}&=&-\sqrt{3}ia_0s^{12}_{1y}{\rm sin}y{\rm sin}x+3a_0s^{12}_{2y}{\rm sin}y {\rm cos}x\nonumber \\
f^{p}_{x13}&=&-3ia_0s^{12}_{1y}{\rm cos}y{\rm cos}x+\sqrt{3}a_0s^{12}_{2y}{\rm cos}y{\rm sin}x\nonumber\\
f^{p}_{x22}&=&-\frac{\sqrt{3}}{2}a_0(s^{22}_{11y}+3s^{22}_{22y}){\rm sin}x{\rm cos}y \nonumber \\
f^{p}_{x23}&=&\frac{3}{2}a_0(s^{22}_{11y}-s^{22}_{22y}){\rm cos}x {\rm sin}y+2\sqrt{3}ia_0s^{22}_{12y}{\rm sin}x {\rm sin}y \nonumber\\
f^{p}_{x33}&=&-\frac{\sqrt{3}}{2}a_0(3s^{22}_{11y}+s^{22}_{22y}){\rm sin}x{\rm cos}y,\label{fpx}
\end{eqnarray}

\begin{eqnarray}
f^{d}_{xx11}&=&-3a^2_0(s^{11}_y+2s^{11}_{yz}{\rm cos}2z){\rm cos}x{\rm cos}y \nonumber\\
f^{d}_{xx12}&=&-\frac{3}{2}ia^2_0s^{12}_{1y}{\rm sin}y{\rm cos}x-\frac{3\sqrt{3}}{2}a^2_0s^{12}_{2y}{\rm sin}y {\rm sin}xx\nonumber \\
f^{d}_{xx13}&=&\frac{3\sqrt{3}}{2}ia^2_0s^{12}_{1y}{\rm cos}y{\rm sin}x+\frac{3}{2}a^2_0s^{12}_{2y}{\rm cos}y{\rm cos}x\nonumber\\
f^{d}_{xx22}&=&-\frac{3}{4}a^2_0(s^{22}_{11y}+3s^{22}_{22y}){\rm cos}x{\rm cos}y \nonumber \\
f^{d}_{xx23}&=&-\frac{3\sqrt{3}}{4}a^2_0(s^{22}_{11y}-s^{22}_{22y}){\rm sin}x {\rm sin}y+3ia^2_0s^{22}_{12y}{\rm cos}x {\rm sin}y \nonumber\\
f^{d}_{xx33}&=&-\frac{3}{4}a^2_0(3s^{22}_{11y}+s^{22}_{22y}){\rm cos}x{\rm cos}y
\end{eqnarray}

\subsection{Superfluid density at low temperature for $p_z$-wave pairing state }
Here we investigate the temperature-dependence of the superfluid density of the $p_z$-wave pairing state in this Q1D system at low temperature $T\ll T_c$. The diamagnetic part of the superfluid density is almost temperature independent at low temperature, and therefore we shall focus on the paramagnetic part, which represents for the consuming of the superfluid density through the nodal quasi-particle excitation near $k_z=0$. At low temperature, only the intra-band quasi-particle excitation is important, which dictates us to simplify the paramagnetic part in Eq.(7) of the main text as,
\begin{eqnarray}
\rho^{xx/yy}_p&=&\frac{1}{2N}\sum_{\mathbf{k}m}|\phi^\dag_{m\mathbf{k}}F^p_{x/y}(\mathbf{k})\phi_{m\mathbf{k}}|^2\frac{\partial n_F(E_{\mathbf{k}m}) }{\partial E_{\mathbf{k}m}}  \label{rouxy}
\end{eqnarray}
\begin{eqnarray}
\rho^{zz}_p&=&\frac{1}{2N}\sum_{\mathbf{k}m}|\phi^\dag_{m\mathbf{k}}F^p_{z}(\mathbf{k})\phi_{m\mathbf{k}}|^2\frac{\partial n_F(E_{\mathbf{k}m}) }{\partial E_{\mathbf{k}m}}.\label{rouz}
\end{eqnarray}
At low temperature $T\ll T_c\approx O(\Delta_{\nu})$, the derivative term $\frac{\partial n_F(E_{\mathbf{k}m}) }{\partial E_{\mathbf{k}m}}=-1/[4k_BT\cosh^{2}(E_{\mathbf{k}m}/2k_BT)]$ is obviously non-zero only for those momenta $\mathbf{k}m$ with the band energy $E_{\mathbf{k}m}$ satisfying $k_BT\gtrsim|E_{\mathbf{k}m}|\approx |\Delta_{\mathbf{k}m}|\propto |k_z|$, which occupy the region near the line gap nodes at $k_z=0$. Thus the momentum summation in Eq.(\ref{rouxy}) and Eq.(\ref{rouz}) is dominantly contributed from the region near $k_z=0$. At that region, from Eq.(8) in the main text and Eq.(\ref{fpz}), Eq.(\ref{fpy}) and Eq.(\ref{fpx}), we can verify that   $\phi^\dag_{m\mathbf{k}}F^p_{x/y}(\mathbf{k})\phi_{m\mathbf{k}}\sim g_1(k_x,k_y)(\alpha-k^2_z)$ and $\phi^\dag_{m\mathbf{k}}F^p_{z}(\mathbf{k})\phi_{m\mathbf{k}}\sim g_2(k_x,k_y)k_z$. With these approximations, the in-plane superfluid density $\rho^{xx/yy}_p$ can be further estimated as,
\begin{eqnarray}
\rho^{xx/yy}_p&\sim&\int^\infty_{-\infty} N(E) \frac{\partial n_F(E) }{\partial E}dE\nonumber\\
&\propto&\int^\infty_{-\infty} |E| \frac{\partial n_F(E) }{\partial E}dE\nonumber\\
&=&-\frac{1}{2T}\int^\infty_0 E \frac{1}{cosh^2(E/2k_BT)}dE \propto -T.
\end{eqnarray}
The out-plane superfluid density $\rho^{zz}_p$ is,
\begin{eqnarray}
\rho^{zz}_p&\sim&\int^\infty_{-\infty} N(E) E^2 \frac{\partial n_F(E) }{\partial E}dE\nonumber\\
&\propto&\int^\infty_{-\infty} |E|^3 \frac{\partial n_F(E) }{\partial E}dE\nonumber\\
&=&-\frac{1}{2T}\int^\infty_0 E^3 \frac{1}{cosh^2(E/2k_BT)}dE \propto -T^3.
\end{eqnarray}
Note that on the above, we have used the relation $N(E)\propto |E|$ for the low energy excitations near the line gap nodes.
\end{bibunit}


\begin{thebibliography}{K2Cr3As3}
\bibitem{Wu2014natcom} W. Wu, J. Cheng, K. Matsubayashi, P. Kong, F. Lin, C. Jin, N. Wang, Y. Uwatoko and J. Luo, Nature. Commun. {\bf 5}, 5508 (2014).

\bibitem{Kotegawa2014} H. Kotegawa, S. Nakahara, H. Tou and H. Sugawara, J. Phys. Soc. Jpn. {\bf 83}, 093702 (2014).

\bibitem{Bao2015} J.-K. Bao, J.-Y. Liu, C.-W. Ma, Z.-H. Meng, Z.-T. Tang, Y.-L. Sun, H.-F. Zhai, H. Jiang, H. Bai, C.-M. Feng, Z.-A. Xu and G.-H. Cao, Phys. Rev. X 5, 011013 (2015).

\bibitem{Tang2015prb} Z. T. Tang, J. K. Bao, Y. Liu, Y. L. Sun, A. Ablimit, H. F. Zhai, H. Jiang, C. M. Feng, Z. A. Xu and G. H. Cao, Phys. Rev. B 91, 020506(R) (2015).

\bibitem{Tang2015scm} Z.-T. Tang, J.-K. Bao, Z. Wang, H. Bai, H. Jiang, Y. Liu, H.-F. Zhai, C.-M. Feng, Z.-A. Xu and G.-H. Cao, Science China Materials, 58(1), 16-10 (2015)
\bibitem{Shen2014} Y. Shen, Q. Wang, Y. Hao, B. Pan, Y. Feng, Q. Huang, L. W. Harriger, J. B. Leao, Y. Zhao, R. M. Chisnell, J. W. Lynn, H. Cao, J. Hu and J. Zhao, arXiv:1409.6615.
\bibitem{Kotegawa2015} H. Kotegawa, H. Tou, H. Sugawara and H. Harima, Phys. Rev. Lett. {\bf 114}, 117002 (2015).

\bibitem{Zhi2015} H. Z. Zhi, T. Imai, F. L. Ning, J.-K. Bao and G.-H. Cao,  Phys. Rev. Lett. {\bf 114}, 14700 (2015).
\bibitem{Pang2015} G. M. Pang, M. Smidman, W. B. Jiang, J. K. Bao, Z. F. Weng, Y. F. Wang, L. Jiao, J. L. Zhang, G. H. Cao and H. Q. Yuan,  arXiv:1501.01880.



\bibitem{Kong2015} Tai Kong, Sergey L. Bud'ko, and Paul C. Canfield, Phys. Rev. B {\bf 91}, 020507 (2015).
\bibitem{Balakirev2015} F. F. Balakirev, T. Kong, M. Jaime, R. D. McDonald, C. H. Mielke, A. Gurevich, P. C. Canfield, S. L. Bud'ko, arXiv:1505.05547 (2015).
\bibitem{Wang2015}    X. F. Wang, C. Roncaioli, C. Eckberg, H. Kim, Y. Nakajima, S. R. Saha, P. Y. Zavalij, J. Paglione, arXiv:1505.07051 (2015).

\bibitem{Zhang2015arxiv} W.-L. Zhang, H. Li, Dai Xia, H. W. Liu, Y.-G. Shi, J. L. Luo, Jiangping Hu, P. Richard, H. Ding. arXiv:1506.01121 (2015).


\bibitem{Jiang2014} H. Jiang, G. Cao and C. Cao,  arXiv:1412.1309.

\bibitem{Wu2015cpl} X. X. Wu, C. C. Le, J. Yuan, H. Fan and J. P. Hu, Chin. Phys. Lett. {\bf 32},057401(2015).

\bibitem{Zhou2015} Y. Zhou, C. Cao and F. C. Zhang, arXiv:1502.03928 (2015).
\bibitem{Wu2015rpa} X. X. Wu, F. Yang, C. C. Le, H. Fan, J. P. Hu, arXiv:1503.06707 (2015).
\bibitem{Zhong2015} H. T. Zhong, X. Y. Feng, H. Chen, J. H. Dai, arXiv:1503.08965 (2015).
\bibitem{M¨¹hlschlegel1959} B. M¨¹hlschlegel, Z. Phys., {\bf 155} 313 (1959).

\bibitem{Luo2015} J. L. Luo (private communication).

\bibitem{Adroja2015} D. T. Adroja, A. Bhattacharyya, M. Telling, Yu. Feng, M.
Smidman, B. Pan, J. Zhao, A. D. Hillier, F. L. Pratt, and A. M. Strydom, arXiv:1505.05743 (2015).


\bibitem{Einzel1986} D. Einzel, P. J. Hirschfeld, F. Gross, B. S. Chandrasekhar, K. Andres, H. R. Ott, J. Beuers, Z. Fisk, and J. L. Smith, Phys. Rev. Lett. {\bf 56}, 2513 (1986).

\bibitem{Gross1986} F. Gross, B. S. Chandrasekhar, D. Einzel, K. Andres, P. J. Hirschfeld, H. R. Ott, J. Beuers, Z. Fisk, J. L. Smith, Z. Phys. B: Condens. Matter, {\bf 64}, 175 (1986).


\bibitem{squid} D. J. Van Harlingen, Rev. Mod. Phys. \textbf{67}, 515 (1995).

\bibitem{Wollman1993} D. A. Wollman, D. J. Van Harlingen, W. C. Lee, D. M. Ginsberg, and A. J. Leggett
Phys. Rev. Lett. {\bf 71}, 213 (1993).

\bibitem{Mathai1995} A. Mathai, Y. Gim, R.C. Black, A. Amar, and F.C. Wellstood, Phys. Rev. Lett. {\bf 74}, 4523 (1995).

\bibitem{josephson}
Y. Asano, Y. Tanaka, M. Sigrist and S. Kashiwaya, Phys. Rev. B \textbf{71}, 214501 (2005).


\bibitem{yang}
L.-D. Zhang, F. Yang and Y. Yao, Sci. Rep. \textbf{5}, 8203 (2015).

\bibitem{Shivaram1986} B. S. Shivaram, T. F. Rosenbaum, and D. G. Hinks, Phys. Rev. Lett. {\bf 57}, 1259 (1986).
\bibitem{Choi1991} C. H. Choi and J. A. Sauls, Phys. Rev. Lett. {\bf 66}, 484 (1991).
\bibitem{Choi1993}  C. H. Choi and J. A. Sauls, Phys. Rev. B {\bf 48}, 1368 (1993).







\end{thebibliography}
\end{document}